\documentclass[
superscriptaddress,
preprint,
amsmath, amssymb,
aps,
prab,
floatfix,
]{revtex4-2}
\usepackage{graphicx,hyperref}
\usepackage{xcolor}
\usepackage[capitalise]{cleveref}
\usepackage{csquotes}

\usepackage[section]{placeins}

\creflabelformat{equation}{#2\textup{#1}#3}
\Crefname{equation}{Eqn.}{Eqns.}
\Crefname{figure}{Fig.}{Figs.}

\graphicspath{{figs}}

\begin{document}

\title{Post-Cold War Diaspora of Russian Particle Physicists}
\author{Vladimir Shiltsev}
\affiliation{Northern Illinois University, DeKalb, IL, 60115, USA}


\begin{abstract}
While the migration of scientists from the Soviet Union to the West occurred at a modest pace during the 1970s–1980s, the dissolution of the USSR in 1991 and the ensuing economic and social hardships precipitated a massive exodus that amounted to a true brain drain. The international physics community—particularly in Europe and the United States—absorbed a substantial influx of specialists in nuclear, high-energy, and accelerator physics, including both seasoned scientists and engineers as well as promising graduate students and postdoctoral fellows. Many of these émigré researchers went on to assume leadership positions, drive major experimental and theoretical initiatives, and achieve scientific distinction that equaled or even surpassed their accomplishments in the USSR/Russia.

In this article we explore the defining features of this post–Cold War scientific diaspora, assess its impact on Russia’s research infrastructure and capabilities, and evaluate its enduring contributions to global particle-physics collaborations and discoveries.
\end{abstract}

\maketitle
\vspace*{-0.02cm} 

\section{Introduction: The Post–Cold War Scientific Exodus}

The post–Cold War “brain drain” from Russia and Former Soviet Union (FSU) ranks among the most consequential scientific migrations of the late twentieth and early twenty–first centuries. Beginning with limited and largely informal outflows in the 1970s–1980s, it accelerated dramatically after the dissolution of the Soviet Union in 1991, reshaping both Russian and international scientific communities. The phenomenon remains deeply ambivalent: widely perceived as a national loss—a massive outflow of human capital and intellectual potential—yet also as a potential long–term resource, as the diaspora continues to engage with home institutions through collaborations, exchanges, and joint projects. For host countries, particularly the United States and those of Western Europe, the influx of Soviet–trained scientists proved a substantial gain, strengthening research capacity, diversifying intellectual traditions, and accelerating the development of entire scientific subfields.

The debate over “brain drain” has persisted in Russia for decades, but within the broader context of globalization and an increasingly integrated global labor market, it has acquired universal significance. The migration of highly skilled professionals—scientists, engineers, and academics—has become a structural feature of the international knowledge economy rather than an isolated national concern. In Russia’s case, the issue was especially acute: the collapse of the Soviet Union unleashed not only massive internal population movements but also extensive emigration of highly qualified specialists, whose departure coincided with the near–collapse of the state research infrastructure.

For centuries, Russia had maintained tightly controlled borders, with external migration either prohibited or heavily restricted. Under Soviet rule, these controls intensified, and the circulation of scientists and intellectuals was confined to officially sanctioned exchanges. Most relocations occurred within the USSR itself—often directed from central regions to new industrial or scientific centers in Siberia, Central Asia, or the Far East.

This rigid system disintegrated abruptly in 1991. The Russian Federation soon became the core of what was, at the time, the world’s second–largest migration system after the United States. For the first time since the 1917 Revolution and the Civil War, Russia experienced large–scale legal emigration of scientists and engineers. In parallel, hundreds of thousands of professionals departed temporarily for research fellowships, sabbaticals, and international collaborations, many of which gradually evolved into permanent relocations.

The post–Cold War Russian scientific diaspora thus illustrates both the vulnerabilities of national research systems under conditions of economic collapse and the enduring internationalism of science itself. While Russia sustained deep institutional and human losses, global science benefited from a remarkable infusion of talent whose influence continues to resonate across disciplines and continents.

\section{The Scientific Diaspora and Its Drivers}

The impact of the post–Soviet crisis on Russian science was profound. During the 1990s and 2000s, roughly half of Russia’s researchers left the scientific profession altogether, seeking employment in business, education, or technical services. Around 45\% managed to remain active in their research fields—often through short-term international contracts, foreign grants, or second affiliations—while an estimated 5\% emigrated permanently. In absolute numbers, this amounted to roughly 15,000–20,000 researchers, distributed primarily across Europe (over 40\%) and the United States (around 30\%), with smaller contingents settling in Israel, FSU countries, Japan, and other parts of Asia. Many others engaged in long-term visiting positions that blurred the boundary between temporary and permanent migration \cite{agamova2007utechka,  florinskaya2018new, korobkov2020russian, subbotin2021brain}.

Physics—and particularly high-energy, nuclear, and accelerator research—was disproportionately represented in this exodus. Approximately 40\% of emigrant scientists came from physics-related fields, including an estimated 2,000–3,000 specialists in particle and nuclear physics, astrophysics, and related technologies. This group accounted for about 5–7\% of the global research community in these disciplines at the time. The presence of USSR/Russian physicists became a defining feature of the global research landscape from the mid–1990s onward, reflecting both the severity of Russia’s scientific crisis and the exceptional training of its pre–existing research schools.

\begin{figure}
    \centering
 \includegraphics[width=0.85\linewidth]{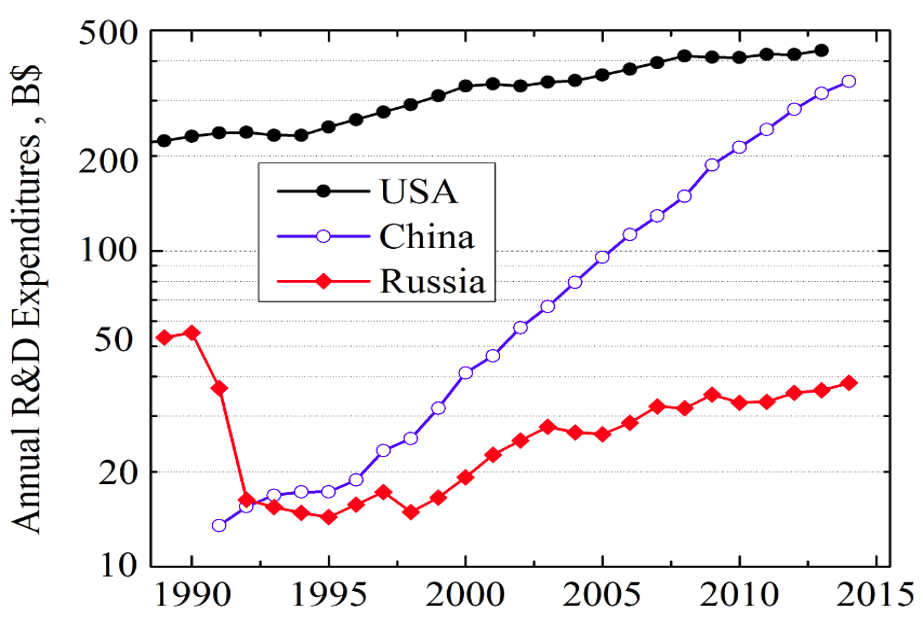}
    \caption{OECD data on annual R\&D expenditures in the United States, China, and Russia, 1989–2014 (in billions of PPP-adjusted U.S. dollars) \cite{oecd}. The vertical axis is shown on a logarithmic scale.}
    \label{fig:gerd}
\end{figure}

The motivations behind this large-scale outflow were multifaceted but interrelated. The economic crisis following the collapse of the USSR was the most immediate cause. The abrupt contraction of research funding (see Fig.\ref{fig:gerd}), the closure or downsizing of entire institutes, and the disappearance of stable academic careers left many scientists without institutional or financial support. Salaries at research centers fell below subsistence levels, forcing many to take on additional work unrelated to their expertise. The prestige once associated with scientific professions eroded rapidly as state priorities shifted toward short-term economic stabilization and privatization.

Equally important was the perceived lack of professional prospects. Laboratories were deprived of modern equipment and access to new technologies. Research themes were often disconnected from global developments, and travel restrictions, though easing, still limited participation in international collaborations. By contrast, opportunities abroad appeared abundant: Western institutions offered advanced facilities, competitive salaries, and the promise of scientific continuity. For many, the choice was not between loyalty and departure, but between professional extinction and survival.

This exodus was further facilitated by several structural and cultural factors. The liberalization of emigration policy in the early 1990s removed many bureaucratic barriers, while the expansion of international collaborations provided natural channels of mobility. Personal connections, developed through conferences and joint experiments, were instrumental in enabling scientists to relocate or obtain temporary contracts abroad. The migration pattern that emerged was often stepwise: short-term visiting appointments were followed by longer fellowships and eventually by permanent settlement, typically at a ratio of about three temporary visits for each long-term emigration.

The highly qualified nature of the emigrant cohort was another defining feature. A large fraction held advanced degrees: roughly 20\% were Doctors of Science (Habilitation level) and more than half were Candidates of Science (PhD equivalent). These levels of expertise allowed rapid integration into host institutions, where Russian-trained physicists became valuable contributors to ongoing projects in both experimental and theoretical research.

Beyond their individual achievements, members of the diaspora played a critical “bridge” role between Russian and Western scientific communities. They maintained professional networks, facilitated access to equipment and publications, co-supervised students, and often mediated joint projects that provided vital lifelines for colleagues who remained in Russia \cite{dezhina2015prospects}. Through these channels, the diaspora helped sustain Russian physics during its most difficult decade, ensuring the continuity of research traditions until domestic reinvestment in science began to recover in the 2010s–2020s.


\section{USSR/FSU/Russian Particle Physics Diaspora}

The roots of the post–Cold War Russian/FSU scientific diaspora in particle physics lie in a long tradition of international collaboration that began well before the dissolution of the Soviet Union. Throughout the 1970s and 1980s, Soviet research institutes maintained formal cooperative links with major laboratories abroad—most notably at CERN, DESY, Fermilab, and KEK—despite the political barriers of the era \cite{pronskikh2016pip, hoddeson2019fermilab, Yarba2022book, pronskikh2025cultural, hof2025competing}. These collaborations were typically mediated through intergovernmental agreements or the Joint Institute for Nuclear Research (JINR, Dubna), which served as the Soviet Union’s official gateway to the global high-energy physics (HEP) community.

\begin{table}[htbp]
\centering
\begin{tabular}{|c|l|c|}
\hline
\textbf{\#} & \textbf{Soviet and Russian Labs and Universities} & \textbf{Number of Experiments} \\
\hline
1  & JINR, Dubna                          & 28 \\
2  & IHEP, Protvino                       & 17 + accelerators \\
3  & ITEP, Moscow                         & 16 \\
4  & Lebedev Physical Institute, Moscow   & 11 \\
5  & PNPI, St. Petersburg                 & 9 + accelerators  \\
6  & Moscow State University              & 8  \\
7  & Institute of Nuclear Research, Moscow & 4  \\
8  & Kiev National University             & 2  \\
9  & Tashkent Physical Technical Institute & 2  \\
10 & Kharkov Physical Technical Institute  & 1  \\
11 & Tomsk Polytechnic Institute           & 1  \\
12 & Yerevan Physics Institute             & 1  \\
13 & Kazakh State University               & 1  \\
14 & Budker Institute of Nuclear Physics   & 1 + accelerators \\
\hline
\end{tabular}
\caption{Soviet and Russian labs and universities participating in Fermilab experiments in 1972-2017 (from \cite{Yarba2017seminar}).}
\label{tab:russian_labs_experiments}
\end{table}

As shown in Table~\ref{tab:russian_labs_experiments}, at least fourteen Soviet and later Russian/FSU institutions participated in Fermilab experiments between 1972 and 2017~\cite{Yarba2017seminar}. The most active contributors included JINR in Dubna (28 experiments), IHEP in Protvino (17 experiments, in addition to work on accelerator development), ITEP and the Lebedev Physical Institute in Moscow (16 and 11, respectively), and the Petersburg Nuclear Physics Institute (PNPI) in St. Petersburg (9 experiments and accelerator projects). Several major universities—Moscow State University, the Institute for Nuclear Research (INR), and others across USSR (Kiev, Tashkent, Kharkov, Tomsk, Yerevan, and Almaty) —also maintained smaller but steady participation. This broad engagement illustrates both the scale and the diversity of Soviet involvement in international high-energy research.

These early collaborative networks became the structural backbone for the subsequent diaspora. When the Soviet Union collapsed and funding for domestic research collapsed with it, many scientists who had participated in joint projects already possessed the professional contacts, reputational capital, and technical experience necessary to transition into permanent or long-term positions abroad. In this sense, the diaspora did not emerge in isolation; it evolved from a well-established matrix of international scientific cooperation.

\begin{table}[!htb]
\centering
\begin{tabular}{|l|c|c c|c|}
\hline
\textbf{Experimental} & \textbf{Number} & \multicolumn{2}{c|}{\textbf{Co-Authors from}} & \textbf{$\quad$ Refs. $\quad$ } \\
\textbf{Collaboration} & \textbf{of Authors} & \textbf{$\quad$ FSU $\quad$} & \textbf{Diaspora $\quad$} &  \\
\hline \hline
\quad  \textbf{FNAL/Tevatron} \quad & & & & {\bf \cite{cdf1995top, d01995top} } \\ 
\quad D0 & 436 & 27 & 15 & \\
\quad CDF & 403 & 0 & 7 &  \\
\textbf{\quad \quad Total} & 839 & 27 & 22 &  \\
\hline
\quad   \textbf{CERN/LEP} $\quad$  & & & & {\bf \cite{lep2013electroweak}} \\
\quad ALEPH & 540 & 0 & 1 & \\
\quad DELPHI & 450 & 17 & 9 &  \\
\quad L3 & 368 & 8 & 20 &  \\
\quad OPAL & 391 & 0 & 1 & \\
\textbf{\quad \quad Total} & 1734 & 25 & 31 & \\
\hline
\quad  \textbf{B-Factories} \quad  & & & & {\bf \cite{belle2001measurement, babar2001measurement} }\\
\quad Belle & 207 & 15 & 1 &  \\
\quad BaBar & 563 & 15 & 8 &  \\
\textbf{\quad \quad Total} & 770 & 30 & 9 &  \\
\hline\hline
\textbf{\quad \quad All expt's} & \textbf{3343} & \textbf{82} & \textbf{62} &  \\
\hline
\end{tabular}
\caption{Examples of FSU and Russian diaspora collaborators in major HEP experiments in 1990s-2000s.}
\label{tab:experiments}
\end{table}

Further insight into the scale and continuity of this outward mobility can be drawn from the participation of scientists from the Former Soviet Union and the Russian diaspora in major international high-energy physics collaborations during the 1990s and 2000s (Table~\ref{tab:experiments}).
Across large experimental efforts such as those at the Tevatron, LEP, and the B-factories, a total of 3,343 co-authors are recorded, of whom 82 were affiliated with institutions in the FSU and 62 represented the diaspora.
The comparable magnitudes of these two groups underscore the strong ongoing engagement of Russian-trained physicists abroad and the persistence of international professional ties that bridge domestic and expatriate segments of the community.
This pattern supports the interpretation that, while Russia experienced a significant net outflow of researchers in the late 1990s and early 2000s, a substantial portion of its scientific workforce continued to contribute actively to global research through institutional affiliations and collaborations outside the country

The experience of large-scale collaborations such as those at Fermilab and CERN also shaped the professional culture of the Russian diaspora. Working within multinational teams accustomed Soviet physicists to open data sharing, distributed responsibilities, and international standards of project management—skills that greatly facilitated their integration into Western laboratories after 1991. Many former collaborators later assumed leadership roles in major experiments at Fermilab, SLAC, DESY, and CERN, extending the legacy of Soviet participation into the era of the Large Hadron Collider and beyond.
%

\begin{figure}
    \centering
    \includegraphics[width=0.85\linewidth]{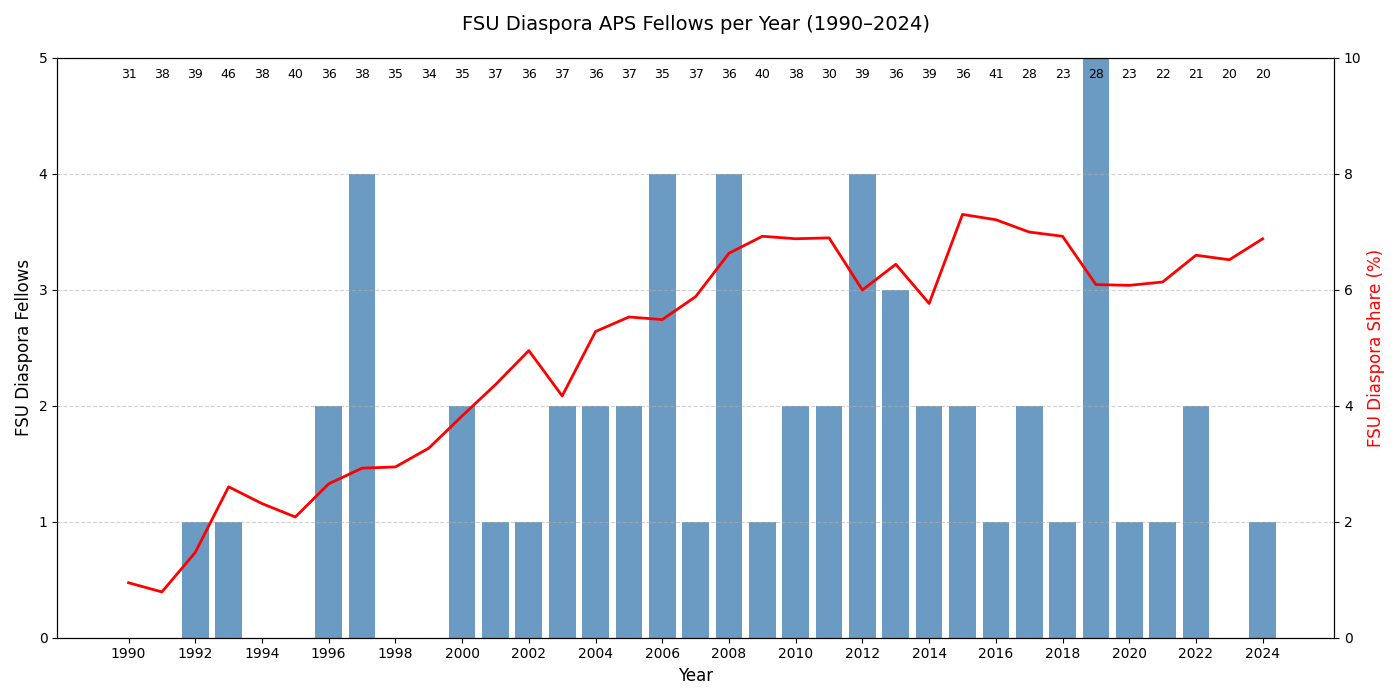}
    \caption{Statistics of APS Fellows elected by DPF, DNP and DPB in 1990-2024.}
    \label{fig:apsfellows}
\end{figure}

The data presented in Figure~\ref{fig:apsfellows} provide an illuminating perspective on the long-term evolution and professional integration of the FSU scientific diaspora in the US particle physics.  The blue bars denote the annual number of newly elected American Physical Society (APS) Fellows of FSU origin in the Divisions of Nuclear Physics (DNP), Particles and Fields (DPF), and Physics of Beams (DPB) combined. 
The figure reveals several notable trends. The early 1990s show a modest but steady rise in the number of FSU-born scientists being recognized as APS Fellows, reflecting the initial wave of migration that followed the dissolution of the USSR and the growing presence of Russian-trained physicists in the U.S. academic and laboratory system. A marked increase occurred during the second half of the 1990s and throughout the 2000s, coinciding with the consolidation of this diaspora into established research groups, particularly in large laboratories such as Fermilab, Brookhaven, SLAC, and several leading universities. During this period, both the absolute number and the relative share of FSU-origin Fellows rose, with the fraction stabilizing near 6-7\% of all newly elected Fellows in the relevant divisions - see the red line that indicates the running average, over 5-year window, of the shere relative to the total number of newly elected Fellows in these divisions (upper line of numbers).

The sustained visibility of the FSU diaspora in APS honors over three decades underscores both their scientific productivity and enduring integration into the international research community. Despite the gradual decline in the overall number of APS Fellowships awarded per year, the diaspora’s representation has remained stable, suggesting continued high impact and leadership in the fields of high-energy, nuclear, and accelerator physics. This long-term pattern mirrors broader trends of international collaboration and scientific mobility, where earlier migration flows have matured into permanent and influential contributions to the host countries’ research ecosystems.

Finally, the persistence of a 6–7\% representation rate suggests that the FSU diaspora has effectively maintained its scientific prominence across generations — from senior scientists who migrated in the 1990s to younger physicists trained abroad but connected through networks of mentorship and collaboration rooted in Soviet scientific schools. The figure thus captures not only a record of recognition but also an enduring legacy of transnational scientific exchange and continuity.

\section{Discussion and Conclusions}

The post–Cold War “brain drain” from Russia, though striking in scale and impact, was neither the first nor the largest wave of scientific migration in modern history. As shown in Table~\ref{table:braindraincomparison}, its magnitude was comparable to earlier transnational flows of researchers—such as those from Europe to the United States during the 1930s and 1940s—but distinctive in its context and structure. It emerged not from political persecution or war, but from a systemic socioeconomic collapse that deeply disrupted the research environment of an entire country.

 \begin{table}[htbp]
\footnotesize 
\centering
\begin{tabular}{|l|c|l|l|l|}
\hline
\textbf{Years} & \textbf{Number of} & \textbf{From} & \textbf{To} & \textbf{Comments} \\ 
\textbf{} & \textbf{Researchers} & \textbf{} & \textbf{} & \textbf{} \\ 
\hline
1720--1800     & $\gtrsim$300           & Switz., Germany, France                    & Russia                 & St.~Petersburg Academy \cite{schulze1985russification, delacroix2021scholars} \\ 
1860--preWWI   & $\gtrsim$3,000        & RUS, E.Europe, IRL       & GER, UK, FRA, USA        & new science centers \cite{bayuk2017, moser2020immigration}\\ 
1930s/40s      & $\gtrsim$3,000         & GER, AUT and satellites            & UK, USA, S.America & Nazi policies \cite{medawar2001hitler} \\ 
1960s--1980s   & $\gtrsim$25,000           & UK, Canada                            & USA                 & {\it “brain drain”}, term coined \\ 
  &           &                             &             & by the Royal Society \cite{royalsoc1963braindrain, royalsoc1987braindrain} \\ 
1970--1980s    & $\sim$3,000            & USSR                               & Israel              & migration of jews \cite{agamova2007utechka}\\ 
1990s--2010s   & $\gtrsim$15,000           & Russia                             & USA, Europe, Asia & USSR break-up \cite{subbotin2021brain, agamova2007utechka} \\ 
2010s--2020s   & $>$30,000           & US, Europe                         & China, Europe       & {\it “reverse brain drain”} \cite{xie2023caught} \\ 
\hline
\end{tabular}
\caption{Examples of historical migration of researchers.}
\label{table:braindraincomparison}
\end{table}

\begin{figure}
    \centering
 \includegraphics[width=0.85\linewidth]{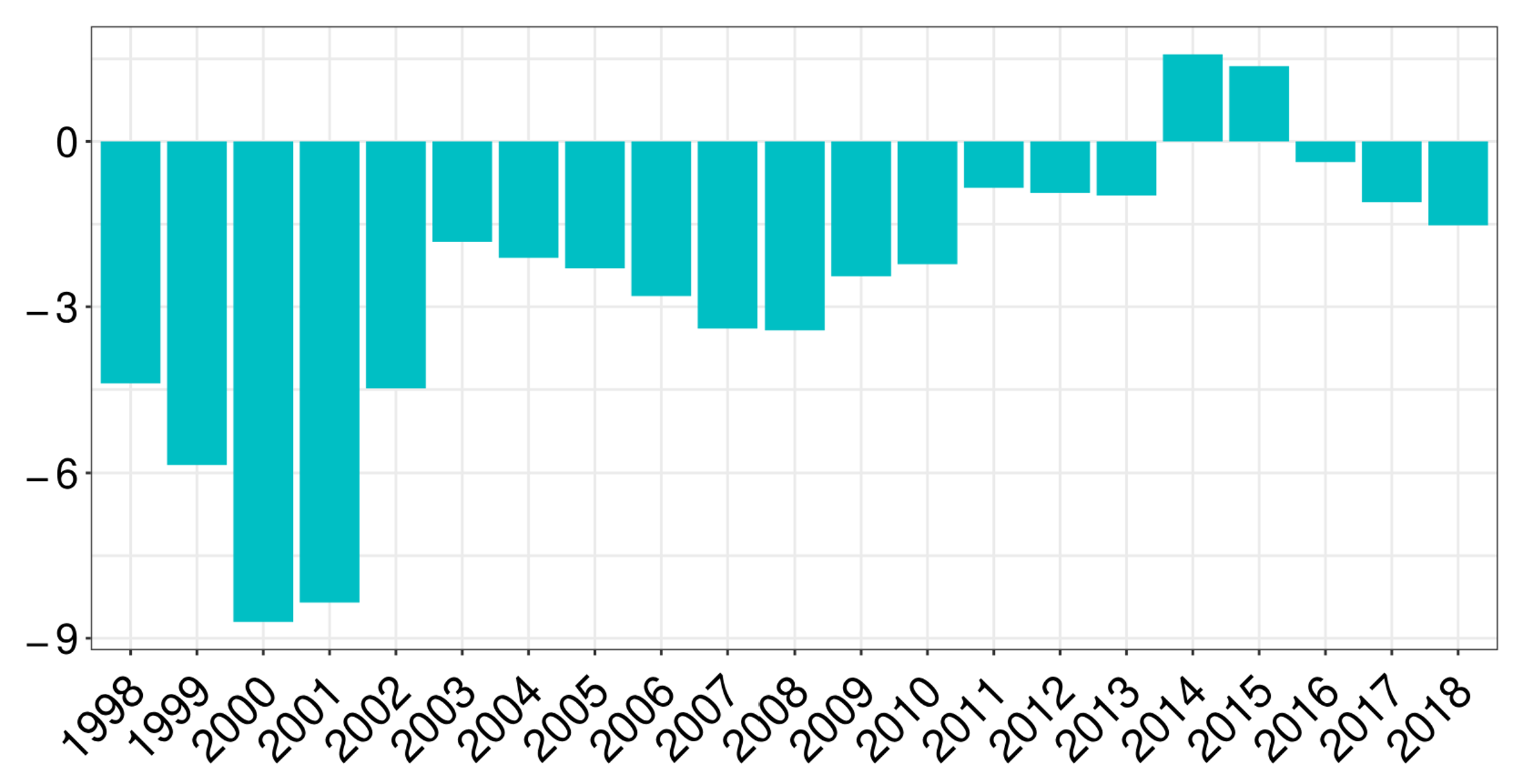}
    \caption{Net migration rates per 1,000 researchers in Russia in 1998-2018. The net flow represents the difference between emigration and immigration, which are estimated to occur in approximately a 3:2 ratio (data from \cite{subbotin2021brain}).
  }
    \label{fig:duration}
\end{figure}

The dynamics of this exodus are illustrated in Figure~\ref{fig:duration}, which traces the duration and intensity of Russian scientific migration over the 1998–2018 period, based on an extensive analysis of more than 2.4 million \textit{Scopus} publications~\cite{subbotin2021brain}.
The migration of researchers with Russian institutional affiliations is inferred from changes in their recorded addresses, which reflect shifts in their countries of affiliation over time.
The lowest net migration rate is observed for the year 2000: in that year, 12.7 per 1,000 published researchers in Russia immigrated to the country (inflow), while 21.4 per 1,000 emigrated (outflow), yielding a negative net balance of 8.7 researchers per 1,000 who left Russia for other countries.
A breakdown of the data by major scientific field indicates that the disciplinary composition of emigrants and immigrants was broadly similar, with researchers in the physical sciences forming the largest group—approximately 45–50\% of the total.

By mid-2010s, the flow had slowed considerably as Russia’s scientific institutions stabilized and international collaboration channels matured. Overall, the emigration wave encompassed on the order of 2,000–3,000 researchers in particle and nuclear physics—approximately (5-7)\% of the world’s workforce in these fields at the time. (Estimates of physics community sizes by subfield are given in Ref.~\cite{battiston2019taking}.)
The principal destinations were Western Europe and North America, with smaller but notable numbers relocating to Israel, Japan, and other parts of Asia. The emigrants included both highly accomplished senior scientists and a large cohort of talented early-career researchers, many of whom went on to build distinguished academic careers and occupy leadership positions in major laboratories and universities. Their integration into host institutions enriched the intellectual diversity and technical capabilities of the global high–energy physics community.

The post-Soviet diaspora in high-energy and nuclear physics was not an abrupt rupture but rather an evolution of pre-existing cooperative frameworks. The long-standing presence of Soviet institutions in international collaborations ensured that, when domestic conditions in Russia deteriorated, the pathways for outward mobility were already open and active. The result was a remarkably smooth transfer of expertise and talent that profoundly influenced both Russian science and the global HEP community.

Beyond individual careers, the diaspora’s broader contribution lay in strengthening the international scientific fabric. USSR/Russian-trained physicists became essential participants in large-scale experiments, collaborations, and facilities worldwide—from CERN and Fermilab to KEK and DESY—facilitating the transfer of knowledge, methods, and experience between Russian and Western institutions. These interactions helped sustain parts of the Russian research infrastructure through difficult decades and demonstrated the resilience of scientific networks under extreme political and economic stress.

The migration was also marked by remarkable acts of professional solidarity. Western colleagues, laboratories, and funding agencies provided crucial support to displaced scientists through fellowships, visiting positions, and institutional partnerships during the hardships of the 1990s and 2000s. This collaborative spirit not only mitigated the negative consequences of the brain drain but also reaffirmed a fundamental truth: that science, at its best, transcends national boundaries and remains a profoundly international enterprise. \\

\section{Acknowledgments}
The author gratefully acknowledges valuable discussions and insightful input from Michael Riordan, Victor Yarba, Dmitri Denisov, Jerry Blazey, and Andrei Gritsan, whose expertise and perspectives on both the history and practice of high–energy physics have greatly contributed to shaping this work. Their comments helped refine the analysis and ensure the accuracy of several historical and technical details presented in the paper.

\bibliographystyle{apsrev}
\bibliography{refs}

@article{florinskaya2018new,
  title={New wave of intellectual emigration from Russia: motives, channels and mechanisms},
  author={Florinskaya, Yu F and Karachurina, LB},
  journal={Monitoring obshchestvennogo mneniya. Ekonomicheskie i sotsialnye peremeny [Monitoring of Public Opinion: Economic and Social Changes]},
  volume={6},
  pages={183--200},
  year={2018}
}

@misc{oecd,
title={{Gross domestic expenditure on R\&D by sector of performance and type of R\&Ds}}, 
author={{OECD}},
note={{OECD Data Explorer}},
url={https://data-explorer.oecd.org/}
}

@misc{Yarba2017seminar,
title={{45 years of collaboration between Fermilab and USSR/Russian scientists}}, 
author={V.Yarba},
note={{Fermilab Colloquium, Sept. 6, 2017}},
url={https://events.fnal.gov/colloquium/archive-2016-2017/},
year={2017}
}

@article{lep2013electroweak,
  title={Electroweak measurements in electron--positron collisions at W-boson-pair energies at LEP},
  author={{ALEPH collaboration and Delphi Collaboration and L3 Collaboration and OPAL collaboration and LEP Electroweak Working Group and others}},
  journal={Physics Reports},
  volume={532},
  number={4},
  pages={119--244},
  year={2013},
  publisher={Elsevier}
}

@article{belle2001measurement,
  title={Measurement of Branching Fractions for B→ $\pi$ $\pi$, K $\pi$, and KK Decays},
  author={Abe, Kazuo and Adachi, I and Ahn, Byoung Sup and Aihara, H and Akatsu, M and Alimonti, G and Asano, Y and Aso, T and Aulchenko, V and Aushev, T and others},
  journal={Physical Review Letters},
  volume={87},
  number={10},
  pages={101801},
  year={2001},
  publisher={APS}
}

@article{babar2001measurement,
  title={Measurement of the Decays B→ $\varphi$ K and B→ $\varphi$ K},
  author={Aubert, Bernard and Boutigny, D and Gaillard, J-M and Hicheur, A and Karyotakis, Yu and Lees, JP and Robbe, P and Tisserand, V and Palano, Antimo and Chen, GP and others},
  journal={Physical Review Letters},
  volume={87},
  number={15},
  pages={151801},
  year={2001},
  publisher={APS}
}

@article{cdf1995top,
  title={Observation of top quark production in p p collisions with the collider detector at fermilab},
  author={Abe, Fumio and Akimoto, H and Akopian, A and Albrow, MG and Amendolia, SR and Amidei, D and Antos, J and Anway-Wiese, C and Aota, S and Apollinari, G and others},
  journal={Physical Review Letters},
  volume={74},
  number={14},
  pages={2626},
  year={1995},
  publisher={APS}
}

@article{d01995top,
  title={Observation of the top quark},
  author={Abachi, Shahriar and Abbott, B and Abolins, M and Acharya, Bannanje Sripath and Adam, I and Adams, DL and Adams, M and Ahn, S and Aihara, H and Alitti, J and others},
  journal={Physical Review Letters},
  volume={74},
  number={14},
  pages={2632},
  year={1995},
  publisher={APS}
}

@article{pronskikh2016pip,
  title={E-36: The first proto-megascience experiment at NAL},
  author={Pronskikh, Vitaly S},
  journal={Physics in Perspective},
  volume={18},
  number={4},
  pages={357--378},
  year={2016},
  publisher={Springer}
}

@book{hoddeson2019fermilab,
  title={Fermilab: Physics, the frontier, and megascience},
  author={Hoddeson, Lillian and Kolb, Adrienne W and Westfall, Catherine},
  year={2019},
  publisher={University of Chicago Press}
}

@article{agamova2007utechka,
  title={Brain Drain from Russia: Causes and Scale},
  author={Agamova, N.S. and Allakhverdyan, A.G.},
  journal={Russian Journal of Chemistry},
  volume={51},
  number={3},
  pages={108--115},
  year={2007}
}

@misc{royalsoc1963braindrain,
  title={{Emigration of
scientists from the United Kingdom:
Report of a committee appointed
by the Council of the Royal Society.}},
  author={{Royal Society Report}},
  journal={Royal Society: London, UK},
url={https://www.jstor.org/stable/41821578},
  year={1963}
}

@misc{royalsoc1987braindrain,
  title={{Migration of academic staff to and from the  The migration of scientists and engineers to and from the UK}},
  author={{Royal Society Report}},
  journal={Royal Society Report},
url={https://royalsociety.org/news-resources/publications/1987/scientist-engineer-migration/},
  year={1997}
}

@article{pronskikh2025cultural,
  title={Cultural Shifts in High Energy Physics Collaboration from the Cold War to the Present: A Historical and Philosophical Perspective},
  author={Petruhina, Polina S and Pronskikh, Vitaly},
  journal={Minerva},
  volume={63},
  number={1},
  pages={135--154},
  year={2025},
  publisher={Springer}
}

@book{Yarba2022book,
    author = "Yarba, Victor A.",
    title = "{Here and There, or There and Here A Physicist's Recollection: Science in the midst of political turmoil in the USSR and USA}",
    isbn = "979-8-4237-5689-5",
    month = "3",
    year = "2022"
}

@article{subbotin2021brain,
  title={Brain drain and brain gain in Russia: Analyzing international migration of researchers by discipline using Scopus bibliometric data 1996--2020},
  author={Subbotin, Alexander and Aref, Samin},
  journal={Scientometrics},
  volume={126},
  number={9},
  pages={7875--7900},
  year={2021},
  publisher={Springer}
}

@article{hof2025competing,
  title={Competing for Collaboration on Particle Accelerators in the Multipolar Cold War World},
  author={Hof, Barbara and Panoutsopoulos, Grigoris and Neto, Clim{\'e}rio Silva},
  journal={Physics in Perspective},
  pages={1--35},
  year={2025},
  publisher={Springer}
}

@article{dezhina2015prospects,
  title={The Prospects for Participation of Russian Expat Scientists in the Development of Russian Science},
  author={Dezhina, Irina},
  journal={Russian Economic Developments},
  number={2},
  pages={35--37},
  year={2015}
}

@article{battiston2019taking,
  title={Taking census of physics},
  author={Battiston, Federico and Musciotto, Federico and Wang, Dashun and Barab{\'a}si, Albert-L{\'a}szl{\'o} and Szell, Michael and Sinatra, Roberta},
  journal={Nature Reviews Physics},
  volume={1},
  number={1},
  pages={89--97},
  year={2019},
  publisher={Nature Publishing Group UK London}
}

@article{delacroix2021scholars,
  title={Scholars and literati at the Saint Petersburg Academy of Sciences (1724--1800)},
  author={De la Croix, David and Doraghi, Mehrdaad},
  journal={Repertorium eruditorum totius Europae},
  volume={5},
  pages={17--26},
  year={2021}
}

@article{schulze1985russification,
  title={The Russification of the St. Petersburg Academy of Sciences and Arts in the eighteenth century},
  author={Schulze, Ludmilla},
  journal={The British Journal for the History of Science},
  volume={18},
  number={3},
  pages={305--335},
  year={1985},
  publisher={Cambridge University Press}
}

@incollection{korobkov2020russian,
  title={Russian academic diaspora: its scale, dynamics, structural characteristics, and ties to the RF},
  author={Korobkov, Andrei V},
  booktitle={Migration from the Newly Independent States: 25 Years After the Collapse of the USSR},
  pages={299--321},
  year={2020},
  publisher={Springer}
}

@article{xie2023caught,
  title={Caught in the crossfire: Fears of Chinese--American scientists},
  author={Xie, Yu and Lin, Xihong and Li, Ju and He, Qian and Huang, Junming},
  journal={Proceedings of the National Academy of Sciences},
  volume={120},
  number={27},
  pages={e2216248120},
  year={2023},
  publisher={National Academy of Sciences}
}

@book{medawar2001hitler,
  title={Hitler's gift: The true story of the scientists expelled by the Nazi regime},
  author={Medawar, Jean S and Pyke, David},
  year={2001},
  publisher={Arcade Publishing}
}

@book{bayuk2017,
  title={People of the World: Russian Scientists Abroad (in Russian)},
  author={Bayuk, Dmitry, ed.},
  year={2018},
  publisher={Alpina Non-Fiction}
}

@article{moser2020immigration,
  title={Immigration, science, and invention. lessons from the quota acts},
  author={Moser, Petra and San, Shmuel},
  journal={Lessons from the Quota Acts (March 21, 2020)},
  year={2020}
}

\end{document}